\documentclass[10pt]{IEEEtran}
\usepackage{amsmath,amsthm,amssymb}
\usepackage{remark}
\usepackage{boxedminipage}
\usepackage{pgfplotstable}
\usepackage{float}
\usepackage{url}
\def\be{\begin{eqnarray}}
\def\ee{\end{eqnarray}}
\def\bee{\begin{eqnarray*}}
\def\eee{\end{eqnarray*}}
   
\newcommand{\ba}{\begin{align}} 
\newcommand{\ea}{\end{align}}

\usepackage{amsfonts} 
\usepackage{amssymb}

\def\be{\begin{eqnarray}}
\def\ee{\end{eqnarray}}
\def\bee{\begin{eqnarray*}}
\def\eee{\end{eqnarray*}}

\newtheorem{thm}{Theorem}

\def\E{\mathbb{E}}
\def\P{\mathbb{P}}

\def\E{\mathbb{E}}
\def\P{\mathbb{P}}

\def\bmx{\begin{pmatrix}}
\def\emx{\end{pmatrix}}

\begin{document}

\newremark{remark}{Remark}

\parindent0mm
\newtheorem{theorem}{Theorem}
\newtheorem{ex}{Example}
\newcommand{\ignore}[1]{}

\title{On the stability of unverified transactions in a DAG-based Distributed Ledger}

\author{P.~Ferraro,\thanks{Pietro Ferraro and Robert Shorten are with the Imperial College London, Dyson School.}
  C.~King,\thanks{Christopher King is with the Department of Mathematics, Northeastern University, Boston, MA 02115 USA.} and
  R.~Shorten\thanks{$\copyright$ 2020 IEEE.  Personal use of this material is permitted.  Permission from IEEE must be obtained for all other uses, in any current or future media, including reprinting/republishing this material for advertising or promotional purposes, creating new collective works, for resale or redistribution to servers or lists, or reuse of any copyrighted component of this work in other works.}}

\maketitle

\begin{abstract}
Directed Acylic Graphs (DAGs)  are emerging as an attractive alternative to traditional blockchain  architectures 
for distributed ledger technology (DLT).  In particular DAG ledgers with stochastic attachment mechanisms potentially
offer many advantages over blockchain, including scalability and faster transaction speeds.
However, the random nature of the attachment mechanism coupled with
the requirement of protection against double-spending transactions might result in an unstable system in which not all transactions
get eventually validated. Such transactions are said to be orphaned, and will never be validated.  
Our principal contribution is to propose a simple modification to the attachment mechanism 
for the Tangle (the IOTA DAG architecture). This modification ensures
that  all transactions 
are validated in finite time, and preserves essential features of the popular Monte-Carlo selection algorithm. 
In order to demonstrate these results we
derive a fluid approximation for the Tangle (in the limit of infinite arrival rate) and prove that
this fluid model exhibits the desired behavior. We also present simulations which validate the results for finite arrival rates.
\end{abstract}

\section{Introductory remarks}
\label{Sec: Introduction}

{\em Bitcoin}, and its underlying technology {\em Blockchain} \cite{Zheng}-\cite{Puthal}, have recently become a source of great debate and 
controversy in both business and scientific communities.  While this debate continues to rage unabated, applications of DLT technologies have begun to emerge across several domains. Applications in logistics \cite{Feng}, banking \cite{SpringerChina}, and in sharing of data \cite{Olnes}, are just a few of the many successes of the technology.  Roughly speaking, applications of DLT technologies have hitherto focussed on two main areas: (i) as a means of (pseudo) monetary payment; and (ii) for tracking goods and services in a trustworthy manner in complex supply chains.  More recently, new application arenas for DLT have been proposed. For example, DLT would appear to offer advantages in applications that require the orchestration of both humans and machines, in applications that require the orchestration of machines and other machines, and more generally in smart city applications, where the issues of social compliance and the enforcement of social contracts are at the forefront (for example discouraging traffic from
breaking regulations, parking for a limited amount of time in restricted areas, etc.) \cite{ourpaper}.  For these types of applications DLT is seen as an interesting  enabling mechanism for a number of reasons. First, some DLT technologies (e.g. Legicash,\protect\footnote{https://legi.cash/} Byteball,\protect\footnote{https://byteball.org/} IOTA \cite{Popov}) are specifically designed for high frequency micro-trading, and in this arena may be more suitable than systems like PayPal or Visa. For example, sometimes low value transactions will not be processed by vendors  (note: however this is also a problem for some DLT architectures such as Bitcoin). 
Second, systems like PayPal or Visa may require a transaction fee, thus making their use in digital deposit based systems questionable; namely, where the entire value of the token is intended to be returned to a compliant agent.
Third, in principle, DLT tokens are more like cash than other digital forms of payment. More specifically, transactions are pseudo-anonymous\protect\footnote{https://laurencetennant.com/papers/anonymity-iota.pdf} because the encrypted address is less revealing than other forms of digital payments. Card based transactions always leave a trail of what was done and when, and are uniquely associated with an individual, the time and location of the spend, and the transaction item. Thus from a privacy perspective (re. Cambridge Analytica and Facebook\protect\footnote{https://www.bbc.com/news/topics/c81zyn0888lt/facebook-cambridge-analytica-data-scandal}), the use of DLT is much more satisfactory than traditional digital transactions.\newline

Our interest in DLT stems from the question of how to enforce social contracts in a smart city context \cite{cdc}. In our prior work we have used the Tangle \cite{ourpaper} for a number of reasons. First, there are no explicit transaction fees associated with IOTA transactions, second, IOTA is designed to support high-frequency micro transactions, and third, the structure of the Tangle is well suited for a mathematical representation  \textcolor{black}{in which its basic properties can be explored in a rigorous setting.  We believe that the IOTA DAG, and other related DLT technologies such as Blockchain, should be of great interest and use to the control and dynamical systems community for a number of reasons. First, the IOTA DAG gives rise to a dynamic system whose essential properties can be captured by characterising the equilibrium state of an appropriate set of time-delay ODE's and PDE's. Second, control theoretic ideas and concepts can be explored to both control and regulate the fundamental properties of the architecture. Finally, as is discussed in \cite{ourpaper}, DLTs can be used to enable feedback strategies to both nudge citizens towards desired behaviour and to price risk in the context of {\em Smart Cities}. While it is beyond the scope of this present paper to discuss these applications, we note that a fundamental question arising in applications of this type concerns token pricing strategies; see \cite{ourpaper} for more details.}\newline 

In the IOTA Tangle, newly arrived transactions select two previous transactions to validate via a proof of work mechanism. A key part of the Tangle is the method for selecting these transactions. This is the so-called {\em tip selection} procedure. To-date, two methods have been widely discussed for tip selection. The first of these is a random selection algorithm, whereby unvalidated transactions are randomly chosen to be validated. The second algorithm is based on a random walk from the interior of the graph (the DAG) to the unvalidated transactions; this is the Markov Chain Monte Carlo (MCMC) selection method. It has been shown in a number of papers that the first of these methods leaves the DAG potentially vulnerable to {\em double spending attacks} by attackers (whereby the same token can be spent more than once), whereas the second of these algorithms may result in situations whereby some transactions are never validated (orphaned transactions that are simply left behind as the DAG grows \cite{Popov}) but is less vulnerable to a double spending attack. The intuition behind the MCMC method is that dishonest tips are unlikely to be selected by newly arriving transactions as they lie on low probability random paths when the MCMC algorithm is initialised from the interior of the DAG graph. Thus, selecting multiple dishonest tips (necessary for the validation of these transaction) is a very low probability event. On the other hand, this might come at the cost of making the system unstable as the number of unapproved tips diverges in time. Our principal contribution in this note is to propose a new {\em hybrid tip selection algorithm} that resolves the dichotomy of double spend avoidance and orphans and results in a Tangle where all transactions are validated in finite time.  As we shall see, our newly proposed algorithm makes the unverified portion of the Tangle bounded in time ensuring that all transactions get approved, and at the same time preserves the feature that selection of multiple dishonest transactions is unlikely.\newline

\subsection{Specific Contributions} The work presented in this manuscript builds on a prior work \cite{ourpaper} in which mathematical models of the  IOTA Tangle with the random tip selection algorithm were developed and validated.  \textcolor{black}{The contributions of this present paper are as follows.}\newline

\begin{itemize}

\item  \textcolor{black}{We develop fluid models of the IOTA DAG behaviour to include Monte Carlo-inspired tip selection algorithms. This gives rise (in the fluid limit) to delayed partial differential equation (PDE) based approximations of the Tangle. Furthermore, this model shows the same unstable behaviour, predicted by simulations, with the MCMC algorithm.}

\item \textcolor{black}{Given the PDE model, we propose a new tip selection algorithm, called {\em hybrid tip selection algorithm}, and prove that for this algorithm, and this model, the Tangle is bounded in time and thus all transactions are eventually validated.}\newline
\end{itemize}

\textcolor{black}{Note that these results significantly extend the results given in \cite{ourpaper}. First, the results in \cite{ourpaper} only treat the elementary random tip selection algorithm, giving rise to an ODE approximation of the Tangle dynamics. The present paper includes analysis of the (preferred and) more secure MCMC tip section algorithm, and gives rise to a PDE model of the Tangle dynamics. Furthermore, we also propose a new tip section algorithm that gives rise to both a secure DAG, and one which leaves no transactions unapproved. These properties are formalised as a theorem based on a PDE fluid limit. We validate these predictions, and demonstrate the efficacy of the new tip selection procedure, using a precise agent based simulation of the Tangle. } \newline

\subsection{Paper structure} This paper is organised as follows: Section \ref{sec: Tangle} provides a qualitative description of the Tangle and of the double spending attack. Simulations show the asymptotic behaviour of the Tangle when the Markov Chain Monte Carlo algorithm is used. Section \ref{sec: hybrid algorithm} discusses a hybrid selection algorithm and the theorem that represents the main contribution of the paper. Section \ref{sec: fluid limit} describes the mathematical modelling of the fluid limit for the Tangle, based on a set of delayed partial differential equations. Finally, Section \ref{sec: Conclusions and Future Research} summarizes the results of this work and outlines future lines of research.

\section{The Tangle}
\label{sec: Tangle}
\begin{figure}
\includegraphics[width=1\columnwidth]{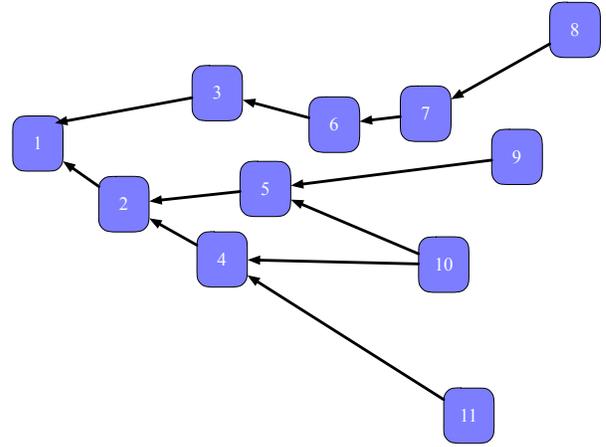}
\caption{Example of a DAG with 11 vertices and 10 edges. All the edges are directed and it is impossible to find a path that connects any vertex with itself. This image was also present in \cite{ourpaper}.}
\label{Fig: DAG}
\end{figure}

In this paper we are interested in a particular DLT architecture that makes use of DAGs to achieve consensus about the shared ledger. A DAG is a finite connected directed graph with no directed cycles. It consists of a finite number of vertices and directed edges, such that there is no directed path that connects a vertex $\nu$ with itself. An example of a DAG is depicted in Figure \ref{Fig: DAG}. \newline

\begin{figure}
\includegraphics[width=1\columnwidth]{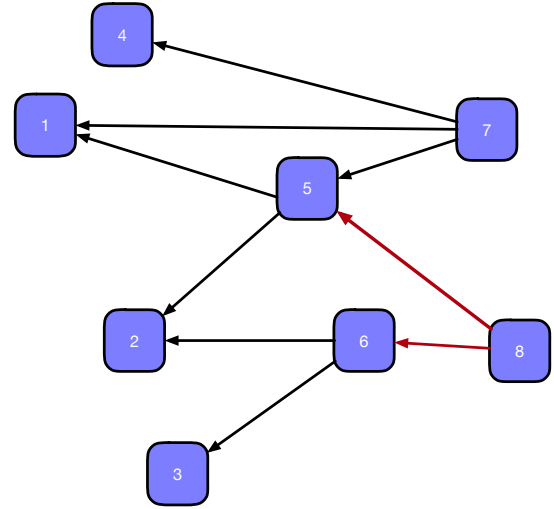}
\caption{Transaction 8 directly approves 5 and 6. It indirectly approves 1, 2 and 3. It does not approve 4 and 7. This image was also present in \cite{ourpaper}.}
\label{Fig: DirUndir Approval}
\end{figure}

\begin{figure}
\includegraphics[width=1\columnwidth]{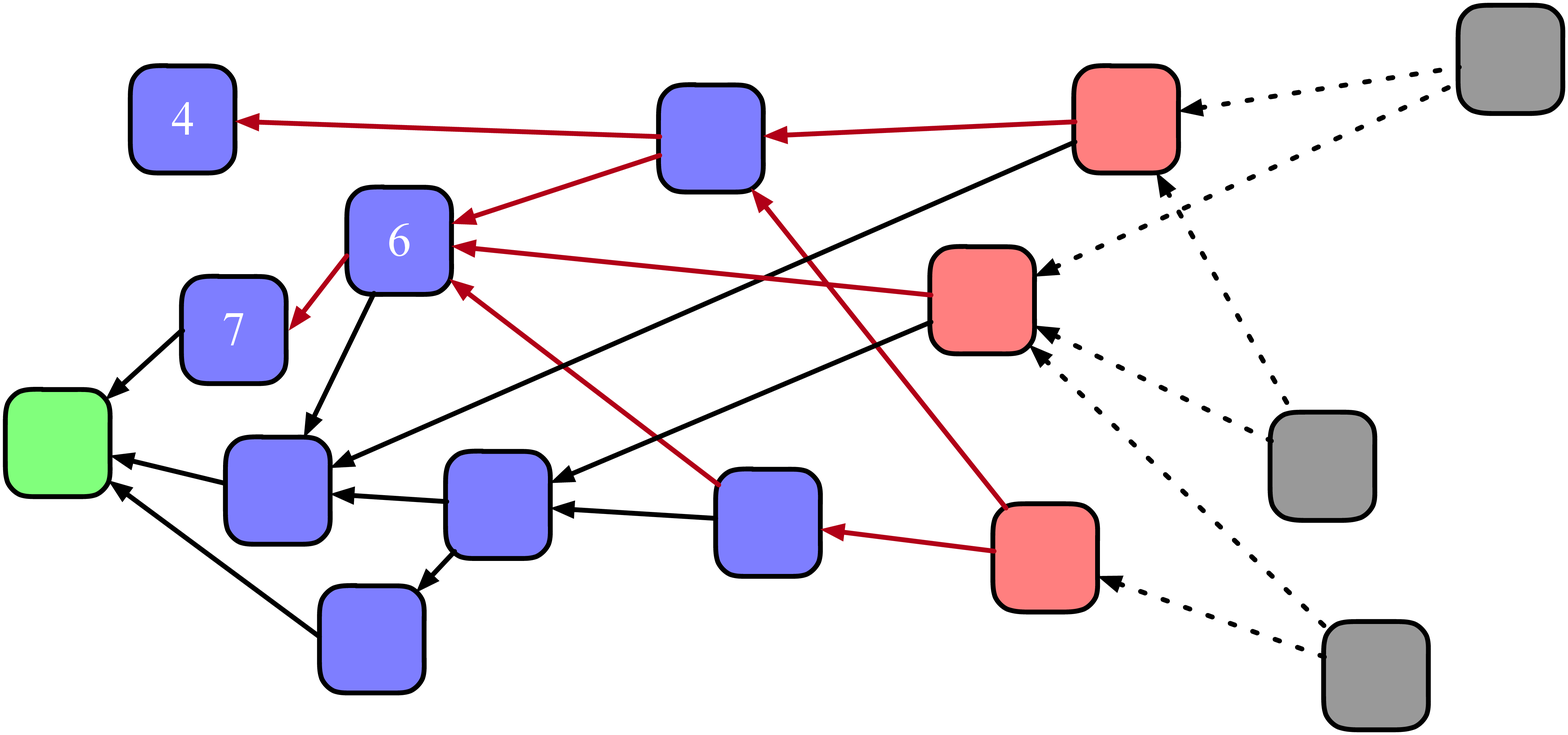}
\includegraphics[width=1\columnwidth]{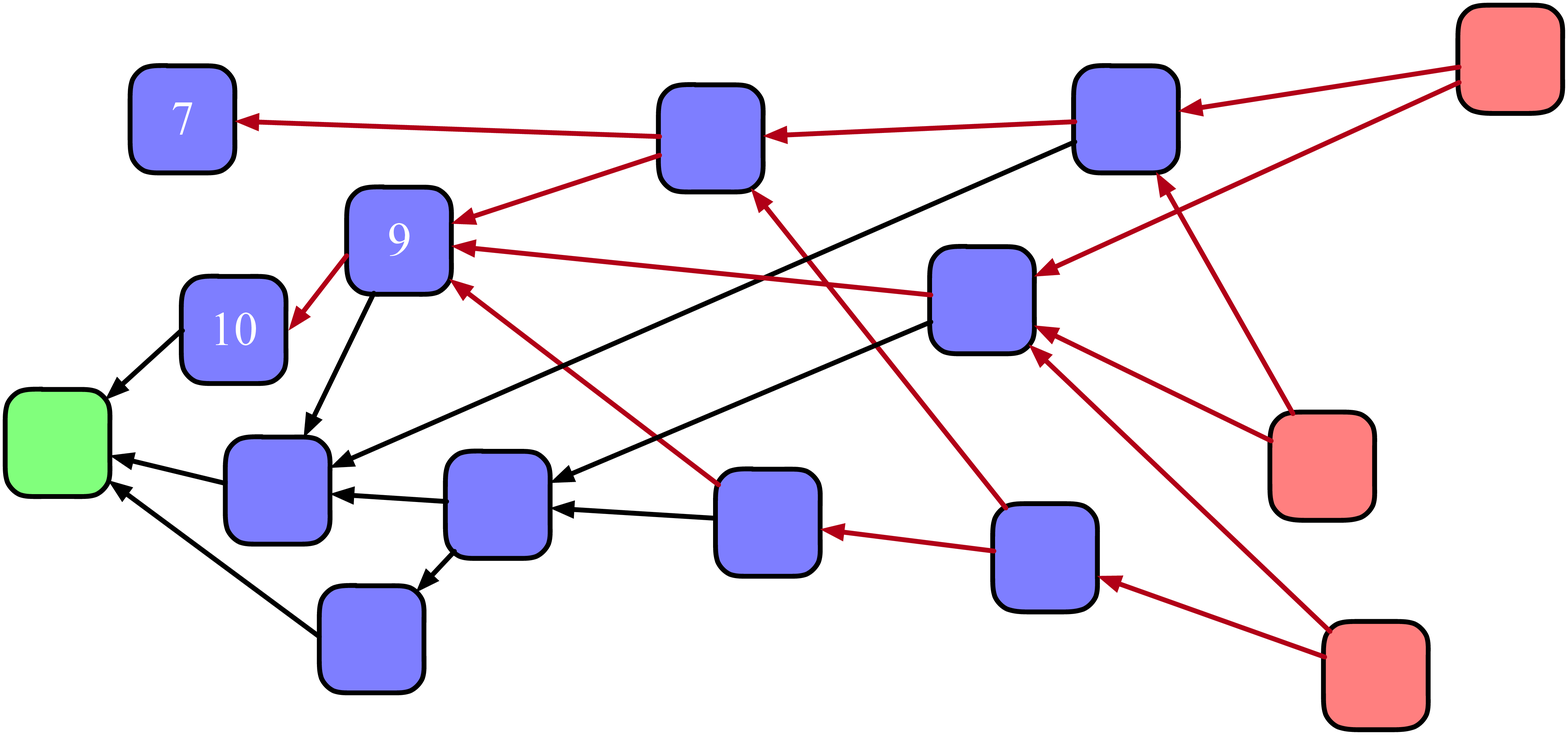}
\caption{Representation of the evolution of the Cumulative weight of three sites as three new transactions enter the Tangle}
\label{Fig: Cumulative weight}
\end{figure}

The Tangle is a particular instance of a DAG based DLT \cite{Popov}. Its objective, according to the original paper, is to provide a cryptocurrency for the IoT industry which has no fees and has low energy consumption. The Tangle is basically a DAG where each vertex or \emph{site} represents a transaction  (we will use interchangably the terms site, transaction and vertex), and 
where the  graph represents the ledger. Before being added to the tangle,  a new transaction must first approve $m$ (normally two) transactions in the tangle. All yet unapproved sites are called \emph{tips} and the set  of all unapproved transactions is called the \emph{tips set}. New transactions will select sites from the tips set for  approval (they are not obliged to do so but it is reasonable to expect them to), although since the approval process takes some time (see comments below concerning Proof of Work), these sites may no longer be tips when the transaction is added to the tangle.
Each successful approval is represented by an edge of the graph, where
a directed edge from site $i$ to site $j$ means that $i$ directly approves $j$. If there is a directed path (but not a single edge) from $i$ to $j$ we say that  $j$ is indirectly approved by $i$ (e.g.,  see Figure \ref{Fig: DirUndir Approval}). The core metric of the Tangle is the \emph{Cumulative weight} of a transaction: this value represents the total number of vertices that approve, directly or indirectly, a given site. Figure \ref{Fig: Cumulative weight} shows an example of how the Cumulative Weight changes in time. The first transaction in the Tangle is called the \emph{genesis} site (namely, the transaction where all the tokens were sent from the original account to all the other accounts) and all transactions indirectly approve it (and therefore its cumulative weight, at every time, is equal to the number of sites in the Tangle). Furthermore, in order to prevent malicious users from spamming the network, the approval step requires a Proof of Work (PoW). This step is less computationally intense than its Blockchain counterpart  \cite{Blockchain Security 1} - \cite{Blockchain Security 4}, and can be easily carried out by common IoT devices (e.g., smartphones, smart appliances, etc.). As mentioned above the Proof of Work introduces a delay for new transactions before they are added to the tangle.\newline

In what follows, we assume that there is a simple way to verify whether the tips selected for approval by a new transaction are consistent with each other and with all the sites directly or indirectly approved by them (this verification occurs during the approval step). If verification fails, the selection process must be re-run until a set of consistent transactions is found. This consistency property is needed in order to prevent malicious users from tampering with the ledger by means of a \emph{double spending} attack (this is described in more detail in the next Section).\newline

\begin{figure}
\centering
\includegraphics[width=1\columnwidth]{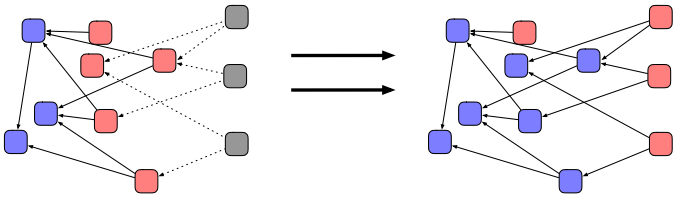}
\caption{Sequence to issue a new transaction. The blue sites represent the approved transactions, the red ones represent the tips and the gray ones represent newly arriving transactions. The black edges represent approvals, whereas the dashed ones represent transactions that are performing the PoW in order to approve two tips. After completing the PoW approved transactions cease to be tips.}
\label{Fig: Tangle}
\end{figure}

As a final note, to illustrate the time evolution of the Tangle, Figure \ref{Fig: Tangle} shows an instance of the Tangle with three new incoming sites (upper panel). The green block (the leftmost) is the genesis transaction, blue blocks are transactions that have already been approved, red blocks represent the current tips of the Tangle and grey blocks are new incoming vertices. Immediately after being  issued,  a new transaction tries to attach itself to $m$ (in this instance two) of the network tips (middle panel). If any of the selected tips was inconsistent with the previous transactions, or with each other, the selection would be rejected and the process would be performed again, until two consistent sites are found. Notice that at this stage, the newly arrived transactions are carrying out the required PoW, and that the tips remain unconfirmed (dashed lines) until this process is over. Once the PoW is finished, the selected tips become confirmed sites and the grey blocks are added to the tips set (lower panel).



\subsection{Double Spending Attack and Tips Selection Algorithms}
\begin{figure}
\includegraphics[width=1\columnwidth]{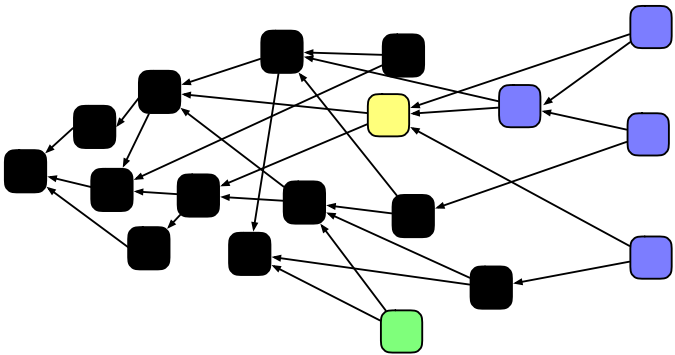}

\caption{The blue and the green transactions are incompatible with each other.}
\label{Fig: DoubleSpending}
\end{figure}

\begin{figure*}[tb!]
\includegraphics[width=1\textwidth]{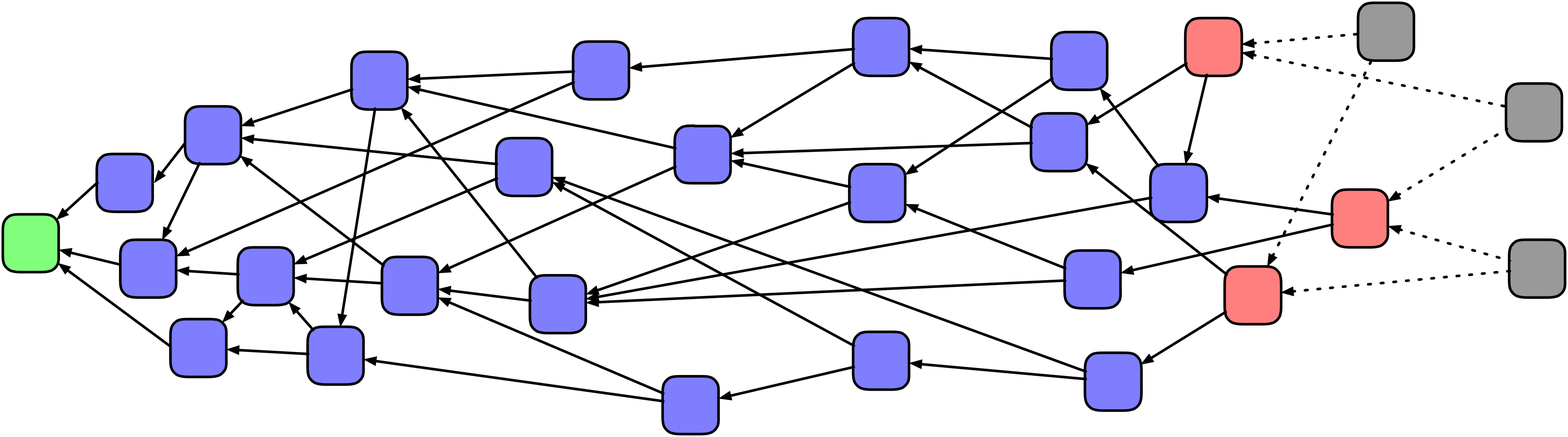}
\includegraphics[width=1\textwidth]{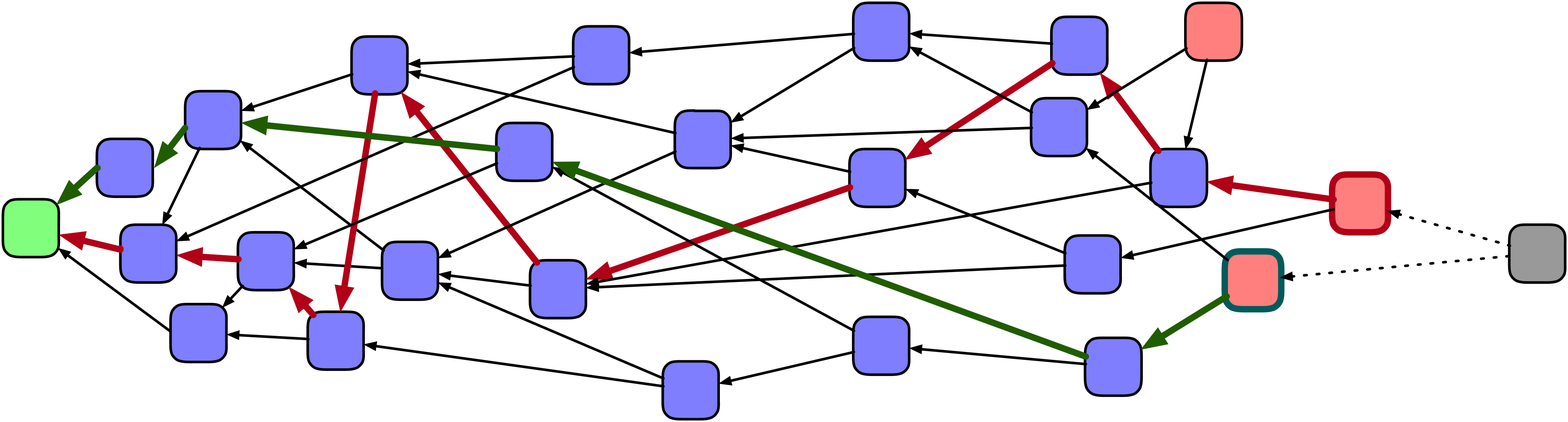}
\caption{Representation of the two main tips selection algorithms: the upper panel shows an instance of the Random Selection algorithm whereas the lower one presents a possible example of the Markov Chain Monte Carlo algorithm}
\label{Fig: Selection Algorithms}
\end{figure*}

A {\em double spending} attack is an attempt to exploit the structure of the DLT in order to spend the same digital token multiple times.\newline 

To see how a { double spending} attack might be carried out on the Tangle, let us consider a specific example to explain with more detail the process of approval. Figure \ref{Fig: DoubleSpending} shows an instance of the Tangle. A malicious user sent a certain amount of currency to a merchant. The corresponding transaction is the yellow site. The same user, afterwards, produces other transactions, where she spends again the same tokens that were sent to the merchant. These correspond to the green blocks. It is worth mentioning, at this point, that there is no mechanism to force a user to select certain sites for approval: any transaction can be considered for selection. Nevertheless it is reasonable to assume that the vast majority of users would have little interest in targeting specific sites for approval and would instead follow the tips selection algorithm that the main protocol proposes. In this scenario, all the transactions that approve the original yellow site (the blue blocks) are incompatible with the green ones, therefore any new transactions can either approve the green/black sites or the blue/black ones. The green/blue combination would be considered invalid (as there is an inconsistency in the ledger) and a new selection would be made. The objective of a hypothetical attacker would be then to wait for the merchant to accept their payment, receive their goods, then create one or more double spending transactions that get approved by other honest sites. It is not possible for two conflicting branches of the Tangle to both continue growing indefinitely (for a formal proof on the subject see \cite{ourpaper}).
Therefore it follows that, if the attacker succeeds in their attempt, the main Tangle will continue to grow from the illegitimate double spending
transaction, and the legitimate branch with the original payment to
the merchant will be \emph{orphaned}, meaning that its sites will not be approved by new transactions. The success probability of such an attack depends on the particular selection algorithm employed and, as mentioned in the previous Section, two algorithms have been proposed so far:\newline

\begin{itemize}
\item \emph{Random Selection Algorithm: } The Random Selection (RS) algorithm selects $m$ (generally two) tips randomly from the pool of all possible tips. The upper panel of Figure \ref{Fig: Selection Algorithms} shows an illustrative example of this procedure. This algorithm, due to its simplicity, makes the Tangle vulnerable to double spending attacks. The interested reader can refer to \cite{Popov} for a detailed discussion on this topic. \newline
\item \emph{Markov Chain Monte Carlo  Algorithm: } The Markov Chain Monte Carlo Selection Algorithm (MCMC) works in a slightly more elaborate way than its random counterpart. In the MCMC algorithm $m$ (generally two) independent random walks are created on the tangle; the walks start at the genesis transaction and move along edges of the tangle. The jumping probability from transaction $j$ to transaction $k$ is proportional to $f(- \alpha(\vartheta_j-\vartheta_k))$, where $f(\cdot)$ is a monotonic increasing function (generally an exponential), $\alpha$ is a positive constant and $\vartheta_i$ represents the Cumulative Weight of transaction $i$. The jumping process stops when the particle reaches a tip, which is then selected for approval. The lower panel of Figure \ref{Fig: Selection Algorithms} shows an example of the paths of two walks in this selection procedure. The main difference with the RS algorithm lies in the use of the graph structure: an attacker would need to create enough transactions, with cumulative weights equal to or larger than the cumulative weight of the main DAG, in order to make the double spending successful. This, in turn, would require the malicious user to possess an amount of computational power comparable to the network of honest users. The interested reader can find more details on this topic in \cite{Popov} \cite{Tangle}.  \newline 
\end{itemize}

The goal of the MCMC algorithm is to increase (and ultimately ensure) the security of the Tangle against double spending attacks performed by malicious users. The algorithm and its security from attacks can be tuned using the parameter $\alpha$: high values of $\alpha$ increase the probability that the particle will jump to the transactions with the largest available cumulative weight (as $\alpha$ approaches $\infty$ the particles will move in a deterministic way), whereas lower values of $\alpha$ make the algorithm and its output more unpredictable (as the cumulative weights matter less and the jumping probability tends to become uniform as $\alpha$ approaches zero). A good way of picturing the effects of this parameter is by comparing it to the inverse temperature of a gas: the smaller the value of $\alpha$ the warmer the gas and vice-versa.\newline

It would seem intuitive that in order to ensure the security of the Tangle the best strategy would be to use the MCMC algorithm with a very large 
value for $\alpha$, 
leading to approval for only the most reliable tips (since the ones at the end of the most likely paths would be chosen by the MCMC walk). Unfortunately, this approach would also mean that a very large number of honest tips would never be approved: indeed large $\alpha$ would imply that newly arrived transactions would select with higher probability a small group of tips (the ones which can be reached along paths where the cumulative weight decreases slowly between successive sites on the path). As time passes an unselected tip would be even less likely to be selected since its distance from the genesis would not increase, while the 
cumulative weight of sites along a path connecting it to the genesis would increase. This outcome is undesirable because it implies that the majority of the transactions would never get approved. However if  small values of $\alpha$ are used, so that most honest tips get eventually approved, then the security against double spending attacks would be compromised. To the best of the authors' knowledge, the only solution proposed so far to this issue is to find a value of $\alpha$ that represents a compromise between the system throughput and its security \cite{Tangle}. To verify that these behaviours do in fact occur we use an agent based simulation of the Tangle: at each time step a random number of transactions arrive, and for each one of these transactions the tip selection algorithm (the MCMC in this scenario) is performed on the current tips set in order to generate graph structures equivalent to the ones presented in detail in Section \ref{sec: Tangle}. In other words, this agent based model simulates the behaviour of each transaction, therefore providing an accurate replica of the mechanism described in Section \ref{sec: Tangle}. We suppose that at every time step the number of new transactions is generated according to a Poisson process with parameter $\lambda$ and that the PoW for the approval procedure of a tip takes $h$ time steps to complete. The variable of interest is the number of tips $L(t)$, which is the number of transactions present in the tips set at the end of each time step.
Due to the stochastic nature of the Tangle, 50 Monte Carlo simulations are performed in order to obtain meaningful results. 

\begin{figure}
\includegraphics[width=1\columnwidth]{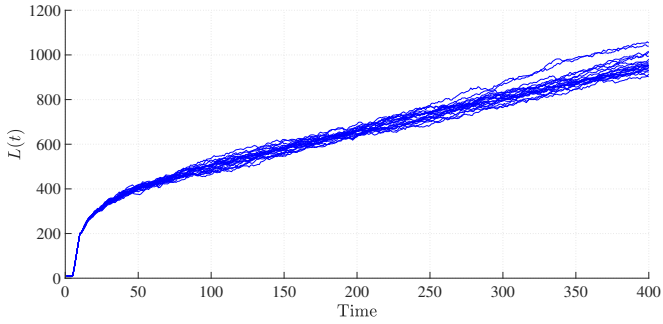}
\caption{50 realizations of the Tangle with the MCMC selection algorithm. The system's parameter are $\lambda = 30$, $h = 5$, $\alpha = 0.1$}
\label{Fig: MCMC unstable}
\end{figure}

\begin{figure}
\includegraphics[width=1\columnwidth]{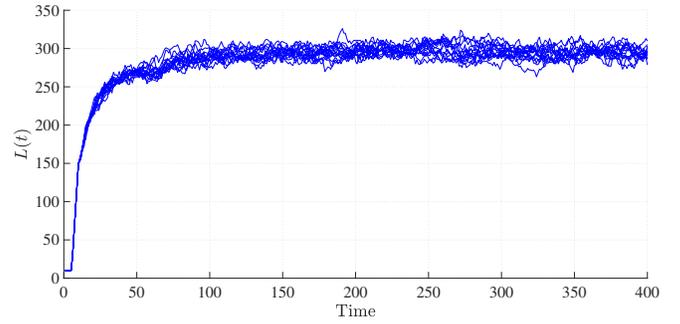}
\caption{50 realizations of the Tangle with the MCMC selection algorithm. The system's parameter are $\lambda = 30$, $h = 5$, $\alpha = 0.001$}
\label{Fig: MCMC stable}
\end{figure}

\medskip
Accordingly, Figures  \ref{Fig: MCMC unstable} and \ref{Fig: MCMC stable} show that for $\alpha = 0.01$ the number of tips diverges with time, whereas for $\alpha = 0.0001$ it fluctuates around a constant value. Clearly, these are only two examples of the possible instances of the Tangle as $\alpha$ varies. From a general point of view, it is very hard to find, in advance, a good trade-off value to make the Tangle simultaneously stable and secure. The Tangle's behaviour depends heavily on parameters such as the rate of arrivals of new transactions and  the time needed to complete the PoW.\newline

 \textcolor{black} {Note that while the Tangle is a stochastic system, its mean field behaviour, for a large number of transactions, is approximately deterministic (refer to the next section or \cite{ourpaper}). As $\alpha$ increases, we observe that the probability that a typical tip is left behind will also increase, and eventually the system will become unstable, resulting in an infinite number of orphaned transactions. On the other hand, while it is reasonably easy to find small values for 
 $\alpha$ such that no transactions remain unverified (e.g., $\alpha = 0$), there is no guarantee that the chosen value of $\alpha$ will result in an acceptable level of security.} \newline

\section{MCMC-like tip selection without orphans}
\label{sec: hybrid algorithm}
As these simulations of the MCMC algorithm illustrate, there is a tension between the need to protect the Tangle against double spending attacks, and the need to ensure that all
transactions eventually get approved. For high values of $\alpha$ the MCMC algorithm favours longer paths from the starting site to the terminal tip, and so
the probability of selecting older tips decreases with time, as the paths to these older tips do not grow. Hence the number of unapproved tips continues to grow, and with time
the probability of selection for older tips decreases to the point that it becomes impossible for old tips to be selected. By contrast small values of $\alpha$ allow older tips to be selected,
however at the same time leaving the Tangle vulnerable to double spending attacks.
This motivates the search for new tip selection algorithms that are not easily exploitable and for which all tips eventually get approved.
In this section we propose one such new algorithm, which we call the {\em hybrid selection algorithm}.\newline 

In order to combine the best properties of the two scenarios (large and small $\alpha$), we propose the following hybrid selection algorithm, which can be divided conceptually into two steps:\newline

\begin{itemize}
\item \emph{Security Step:} the first set of selections is made using the MCMC algorithm with a high value of $\alpha$. This is done to protect security by ensuring that honest tips get selected preferentially;\newline
\item \emph{Swipe Step:} the set of second selections is performed using a different algorithm: it can be a RS or a MCMC with a low value of $\alpha$. This step serves the purpose of taking care of the older transactions that are not likely to be selected by the first step.\newline
\end{itemize}

Roughly speaking, the Security Step takes care of the security of the system, whereas the purpose of the Swipe Step is to ensure that no tips get left behind by the first, more accurate, selection.\newline

\begin{figure}
\includegraphics[width=1\columnwidth]{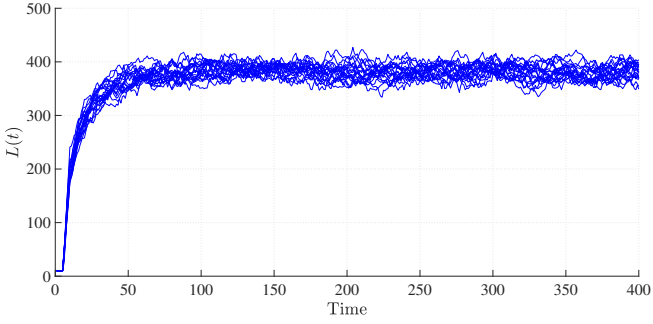}
\caption{50 realizations of the Tangle with the hybrid selection algorithm. The system's parameter are $\lambda = 40$, $h = 5$, $\alpha = 1$}
\label{Fig: Hybrid stable}
\end{figure}

Figure \ref{Fig: Hybrid stable} shows the dynamics of the number of tips $L(t)$ under the hybrid selection algorithm. Note that the values for $\alpha$ and $\lambda$ in this simulation are higher than the unstable simulations performed in Section \ref{sec: Tangle}. The selection procedure, while being as safe from double spending attacks as any Tangle with high values of $\alpha$, also ensures that eventually all tips get approved, thus confirming the theoretical result presented in this section. \newline

In the next section we present a mathematical model for the dynamics of the Tangle under various tip selection algorithms, in the regime of high arrival rate
(also known as the fluid limit). Within this model we formulate and prove a result that demonstrates boundedness of the size of the tip set when the hybrid
algorithm is used.

\section{Mathematics of the DAG: A fluid limit}
\label{sec: fluid limit}
We now present, on the basis of some of the results obtained in \cite{ourpaper}, a mathematical model that approximates the Tangle behaviour with large arrival rate, taking into account the {mechanics} discussed in the previous Section.\newline

The Tangle can be described as an increasing family of DAG's, $\lbrace G(t), t > 0 \rbrace$  where each site of $G(t)$ contains the record of a transaction which arrived at or before time $t$. We are interested in describing the growth of the Tangle $G(t)$ in a situation where one central server keeps the record of all the transactions. In a real network there would be multiple local copies of the Tangle, and each user would independently update its own copy. However due to synchronization issues, the presence of multiple independent servers complicates the analysis; therefore, we assume the simpler scenario with one main server, and multiple users accessing it whenever they want to issue a transaction.\newline

We assume that each newly created transaction selects $m \ge 2$ tips for approval, and attempts to validate them 
(note that $m=2$ in the standard Tangle). If validation fails
the choices are discarded and another
$m$ tips are selected for validation. This continues until the process is successful, and we assume that this whole
validation effort is essentially instantaneous. However after the validation there is a fixed waiting period $h$  during which
the PoW is carried out and the transaction is communicated to the server where the Tangle is stored. 
During this time the approvals of the selected tips are pending, so the tips may still be
available for selection by other new transactions. After the waiting time $h$, new directed edges are
added to the graph, directed from the new site (corresponding to the new transaction) to its $m$ parent sites
(corresponding to the $m$ tips which were successfully validated). After this point the $m$ parent sites
are no longer tips, and so are no longer available for selection by other new transactions.
However it may happen that some of these sites had already ceased  to be tips at an earlier time, due to their being validated by some
other new transaction. 

\medskip
To facilitate the discussion in the sequel, we consider a general algorithm for selecting the tips for validation.
Letting ${\cal L}(t)$ denote the set of tips at time $t$, 
we assume that there are $m$ probability distributions $Q^{(j)}(t) = \{Q^{(j)}_b(t): \, b \in {\cal L}(t)\}$ $(j=1,\dots,m)$
defined on the set of tips at time $t$.  The newly created transaction selects the first tip for approval using
$Q^{(1)}(t)$, selects the second tip using $Q^{(2)}(t)$, and so on to the $m^{th}$ tip. These probability distributions are continually updated as the tangle grows.

\medskip
\noindent The model involves these variables:
\begin{itemize}
\item[1)] ${\cal L}(t)$ is the set of tips at time $t$
\item[2)] ${\cal W}(t)$ is the set of `pending' tips at time $t$ which are being considered for approval by some new transaction
\item[3)] ${\cal X}(t) = {\cal L}(t) \setminus {\cal W}(t)$ is the set of `free' tips at time $t$
\item[4)] $L(t) = | {\cal L}(t) |$, $W(t) = | {\cal W}(t) |$, $X(t) = | {\cal X}(t) |$
\item[5)] $T_a$ is the time when transaction $a$ is created, and the PoW starts
\item[6)] $T_a + h$ is the time when the PoW ends, and the transaction $a$ is added to the tangle
\item[7)] $N(t)$ is the number of transactions created up to time $t$
\item[8)] $|G(t)| = N(t-h)$ is the size of the DAG at time $t$
\item[9)] $U(T_a) \in \{0,1,\dots,m\}$ is the number of free tips selected for approval by transaction $a$ at time $T_a$
\end{itemize}
We have the relations
\be
N(t) & = &  \sum_{a : T_a \le t} \, 1 \label{eqn:N} \\
W(t) &=& \sum_{a: t - h < T_a \le t} U(T_a) \label{eqn:W} \\
X(t) &=& N(t -h) - \sum_{a : T_a \le t} U(T_a) \label{eqn:X} \\
L(t) &=& N(t-h) - \sum_{a : T_a + h \le t} U(T_a) \label{eqn:L}
\ee

\medskip
$U(T_a)$ is a random variable whose distribution depends on the sets
${\cal X}(T_a)$, ${\cal W}(T_a)$ and ${\cal L}(T_a)$, as well as the 
distributions $\{Q^{(j)}(T_a)\}$. (Note: we assume that
variables are updated at the times $\{T_a\}$ and $\{T_a +h\}$ so $U(T_a)$ really depends
on the values just before time $T_a$. For convenience we will ignore this distinction in our notation).
Let $R_b^{(j)}(T_a)$ be the indicator variable of the event that $b$ is the $j^{th}$ tip selected by the new
transaction $a$. Since tips are selected independently, and the same tip may be selected multiple times, we have
\be
U(T_a) =  \sum_{b \in {\cal X}(t)} \, \left[ 1 - \prod_{j=1}^m \left(1 - R_b^{(j)}(T_a) \right) \right]
\ee
It follows that the expected value is
\be\label{exp-U}
\E[U(T_a)] = \sum_{b \in {\cal X}(t)} \, \left[ 1 - \prod_{j=1}^m \left(1 - Q^{(j)}_b(T_a) \right) \right]
\ee

\subsection{Tip selection algorithms}
We will consider a class of selection distributions which are modeled on two well-studied
examples from the literature. The first example is the random tip selection algorithm,
where the distribution $Q^{({\rm ran})}$ is uniform over the tip set ${\cal L}(t)$. In this case
\be
Q^{({\rm ran})}_b(t) = \frac{1}{L(t)}
\ee
and it follows that
\be
\P(\mbox{select a free tip at time $t$ with $Q^{({\rm ran})}$})
= \frac{X(t)}{L(t)}
\ee

The second example is the MCMC attachment algorithm which 
selects tips by generating a random walk along paths in the Tangle from the genesis to the tips.
The algorithm is designed to favor paths where the change in accumulated weight at each step is small. This feature tends to favor paths where each site
was approved relatively soon after being added to the Tangle: if a tip had to wait a long time before it was approved, then its parent sites in the Tangle
would have been more likely to accumulate a higher weight while the tip was waiting, which would have led to a larger change in accumulated weight at the step
from parent to child. Thus the MCMC algorithm will tend to favor tips which are relatively newly added to the Tangle.

\medskip
Because the MCMC algorithm generates a complicated probability distribution on the tips which depends on the detailed structure of the DAG,
we will model it with a simpler class of distribution which we believe captures important features of the MCMC distribution.
We define the {\em age of a tip} to be the time since the tip was added to the Tangle. As argued above, the MCMC algorithm will generate a distribution which is biased
toward tips with smaller age, and thus a suitably age-biased distribution should provide a reasonable model for the behavior of the MCMC algorithm. So we assume that there is a positive function $g(\cdot)$ such that the MCMC distribution $Q^{({\rm MC})}(t)$ has the form
\begin{eqnarray}
Q^{({\rm MC})}_b(t) &=&  \P(\mbox{select tip $b$ at time $t$ with MCMC}) \nonumber \\
&=& Z(t)^{-1} \, g(t - T_b - h) \label{def:Q-g}
\end{eqnarray}
where $T_b+h$ was the time when the tip was first added to the Tangle, so $t - T_b - h$ is the tip's current age, and
where $Z(t)$ is the normalization factor
\be
Z(t) = \sum_{b \in {\cal L}(t)} g(t - T_b - h) 
\ee
We then have
\be
&& \hskip-1in \P(\mbox{select a free tip at time $t$ with $Q^{({\rm MC})}$}) \nonumber \\
&=& \sum_{b \in {\cal X}(t)} \, Z(t)^{-1} \, g(t - T_b - h) \nonumber \\ 
&=&
\frac{\sum_{b \in {\cal X}(t)} \, g(t - T_b - h)}{\sum_{b \in {\cal L}(t)} \,  g(t - T_b - h)}
\ee
We will consider tip selection algorithms of the form (\ref{def:Q-g}) where $g$ is a general positive function.
So we consider functions $g_1,\dots,g_m$, and following (\ref{def:Q-g}) define for $j=1,\dots,m$
\be
Q^{(j)}_b(t) &=& Z_j(t)^{-1} \, g_j(t - T_b - h), \label{def:Q-j-g} \\
 Z_j(t) &=& \sum_{b \in {\cal L}(t)} g_j(t - T_b - h) \label{def:Z-j}
\ee
This class encompasses the random tip selection algorithm (in which case $g$ is constant) as well as the proxy model
for the MCMC algorithm. In the case of the MCMC algorithm the weighting provided by the age-biased factor $g(\cdot)$
will concentrate the distribution on recent arrivals, and thus we will assume that, for MCMC-like distributions, {for at least for some values of the parameter $\alpha$,}
the function $g$ is integrable, that is
\be\label{g-int}
\int_{0}^{\infty} g(s) \, d s < \infty
\ee
Note that the condition (\ref{g-int}) will not hold true for the random tip selection algorithm, where $g$ is constant.

\subsection{The fluid model}
We consider the regime of high arrival rate, where the growth of the Tangle is described by a fluid model.
We will present some plausible arguments to describe the dynamics of the tip set at high arrival rate,
and use these to derive the fluid limit.
We assume a Poisson arrival process with rate $\lambda$, so that the fluid limit corresponds to $\lambda \rightarrow \infty$.
We rescale variables by $\lambda^{-1}$ and define the rescaled numbers of tips to be
\be
l(t) = \lim_{\lambda \rightarrow \infty} \lambda^{-1} \, L(t), \\
x(t) = \lim_{\lambda \rightarrow \infty} \lambda^{-1} \, X(t), \\
w(t) = \lim_{\lambda \rightarrow \infty} \lambda^{-1} \, W(t)
\ee
For a fixed age $s > 0$, the number of tips at time $t$ with ages between $s$ and $s + \delta$ is
\be
\# \{ a \in {\cal L}(t): s \le t - T_a - h \le s + \delta \}
\ee
In the fluid limit we assume that this quantity grows proportionally to $\lambda \, \delta$, so we define the density $l(t,s)$ as
\be
l(t,s) &=& \nonumber \\
&& \hskip -1in 
\lim_{\stackrel{\lambda \rightarrow \infty}{\delta \rightarrow 0}} \left(\lambda \, \delta\right)^{-1} \, 
\# \{ a \in {\cal L}(t): s \le t - T_a - h \le s + \delta \}
\ee
Then the total number of tips at time $t$ (after re-scaling by $\lambda^{-1}$) is
\be
l(t) = \int_{0}^{t-h} l(t,s) \, ds
\ee
(note that $s \le t-h$ since we assume that no tips arrived before time $h$).
We make similar assumptions about the numbers of pending tips and free tips, and define age-dependent densities
$w(t,s)$ and $x(t,s)$ in a similar way.

\medskip
Note that each new transaction selects $m$ tips for approval, and some of these
may not be free (due to having been selected by earlier transactions). Therefore
at time $t$ the number of pending tips $W(t)$ is at most $m$ times the number of new transactions
created in the interval $[t-h,t]$. In the fluid limit this implies the bound
\be\label{bound-w}
w(t) \le m \, h
\ee

\medskip
Over a small time interval $(t, t+ \epsilon)$ the number of arrivals will be $\lambda \, \epsilon$, and these arrivals will
select tips for approval according to the distributions $\{Q^{(j)}(t)\}$ defined in (\ref{def:Q-j-g}).
Using (\ref{exp-U}), it follows that
the expected number of tips selected by this group of arrivals which lie in the age range
$s$ to $s + \epsilon$ is approximately
\be
\lambda \, \epsilon \hskip-0.2in  \sum_{\stackrel{b \in {\cal L}(t):}{s \le t - T_b - h \le s + \epsilon}} \, 
\left[ 1 - \prod_{j=1}^m \left(1 - Q^{(j)}_b(t) \right) \right]
\ee
Similarly the expected number of free tips in the age range $s$ to $s + \epsilon$ selected by this group of arrivals is
\be\label{select-free}
\lambda \, \epsilon \hskip-0.2in  \sum_{\stackrel{b \in {\cal X}(t):}{s \le t - T_b - h \le s + \epsilon}} \, 
\left[ 1 - \prod_{j=1}^m \left(1 - Q^{(j)}_b(t) \right) \right]
\ee
\medskip
Now we consider the evolution of the density of free tips in the age range $(s,s+\epsilon)$
between times $t$ and $t + \epsilon$.
The number of these tips at time $t$ is (approximately) $x(t,s) \epsilon$,
and we will compare this with the number at time $t + \epsilon$.
During this time increment $\epsilon$ the subset of tips of age $(s,s+\epsilon)$ inherits the tips from the 
age range $(s - \epsilon,s)$ while it also loses some tips due to the selection by new arrivals, as detailed above in
(\ref{select-free}). Thus the evolution is
\be\label{x-dyn1}
&& \hskip-0.5in x(t+\epsilon,s) \, \lambda \, \epsilon =  x(t, s - \epsilon) \, \lambda \,\epsilon \nonumber \\
&& \hskip-0.5in -
\lambda \, \epsilon \,
 \sum_{\stackrel{b \in {\cal X}(t):}{s \le t - T_b - h \le s + \epsilon}} \, 
\left[ 1 - \prod_{j=1}^m \left(1 - Q^{(j)}_b(t) \right) \right]
\ee
\normalfont
Furthermore for the $j^{th}$ tip selection we have
\be
Z_j(t) &=& \sum_{b \in {\cal L}(t)} g_j(t - T_b - h) \nonumber \\
& =  & \sum_{k \ge 1} \sum_{\stackrel{b \in {\cal L}(t):}{(k-1) \epsilon \le t - T_b - h < k \epsilon}} g_j(t - T_b - h) \nonumber \\
& \simeq & \sum_{k \ge 1} \, g_j(k \epsilon) \,\lambda \, l(t, k \epsilon) \, \epsilon \nonumber \\
& \simeq & \lambda \, \int_{0}^{t - h} g_j(s) \, l(t,s) \, d s
\ee
We define
\be\label{def:zeta}
\zeta_j(t) =  \int_{0}^{t - h} g_j(s) \, l(t,s) \, d s
\ee
then we also have
\be
&& \sum_{\stackrel{b \in {\cal X}(t):}{s \le t - T_b - h \le s + \epsilon}} \, 
\left[ 1 - \prod_{j=1}^m \left(1 - Q^{(j)}_b(t) \right) \right] \nonumber \\
& \simeq &
\sum_{\stackrel{b \in {\cal X}(t):}{s \le t - T_b - h \le s + \epsilon}} \, 
\left[ 1 - \prod_{j=1}^m \left(1 - \lambda^{-1} \, \frac{g_j(s)}{\zeta_j(t)} \right) \right] \nonumber \\
& \simeq &
\lambda \, x(t,s) \, \epsilon \, 
\left[ 1 - \prod_{j=1}^m \left(1 - \lambda^{-1} \, \frac{g_j(s)}{\zeta_j(t)} \right) \right] \nonumber \\
& = &
\epsilon \, x(t,s) \, \sum_{j=1}^m \frac{g_j(s)}{\zeta_j(t)} + O(\lambda^{-1})
\ee 
Thus in the limit $\lambda \rightarrow \infty$ the evolution equation (\ref{x-dyn1}) becomes
\be\label{x-dyn2}
x(t+\epsilon,s) = x(t, s - \epsilon) - \epsilon \, x(t,s) \, \sum_{j=1}^m \frac{g_j(s)}{\zeta_j(t)}
\ee
and hence in the limit $\epsilon \rightarrow 0$ we get
\be\label{eqnX:1}
\frac{\partial x}{\partial t}(t,s) + \frac{\partial x}{\partial s}(t,s) &=& - x(t,s) \, \sum_{j=1}^m \frac{g_j(s)}{\zeta_j(t)}
\ee
\medskip
This delay equation holds for $s > 0$. At $s=0$ the number of free tips is continually replenished at the rate $\lambda$, as transactions complete the approval process.
Hence (\ref{eqnX:1}) must be supplemented with the boundary condition
\be\label{eqnX:1bc}
x(t,0) = 1.
\ee

\medskip
The dynamical equation for the pending tips can be derived in a similar way. The density $w(t,s)$ receives the outflow from $x(t,s)$, so for $0 \le s \le h$ we have
\be\label{eqnW:1}
\frac{\partial w}{\partial t}(t,s) + \frac{\partial w}{\partial s}(t,s) =  x(t,s) \, \sum_{j=1}^m \frac{g_j(s)}{\zeta_j(t)}
\ee
For $s >h$ there is also an outflow from $w$ due to approval by new transactions which were created at
time $t-h$ and which selected some of these tips for validation, so for $s > h$ we have
\be\label{eqnW:2}
\frac{\partial w}{\partial t}(t,s) + \frac{\partial w}{\partial s}(t,s) &=&  x(t,s) \, \sum_{j=1}^m \frac{g_j(s)}{\zeta_j(t)}  \nonumber \\
&& \hskip-1in - x(t-h,s-h) \, \sum_{j=1}^m \frac{g_j(s-h)}{\zeta_j(t-h)}
\ee
Define
\be\label{def:theta}
\theta(u) = \begin{cases} 0 & \mbox{for $u < 0$} \\ 1 & \mbox{for $u \ge 0$} \end{cases}
\ee
Noting that $l(t,s) = x(t,s) + w(t,s)$ we get
\be\label{eqnL:1}
\frac{\partial l}{\partial t}(t,s) + \frac{\partial l}{\partial s}(t,s) = \nonumber \ee
\be
- x(t-h,s-h) \, \sum_{j=1}^m \frac{g_j(s-h)}{\zeta_j(t-h)} \, \theta(s-h)
\ee
\normalfont
and again this is supplemented by the initial condition
\be\label{eqnL:1bc}
l(t,0) = 1.
\ee

\medskip
In summary: we have the following system of dynamical equations for the tip densities in the fluid limit:
{\small
\be\label{sys1}
\frac{\partial x}{\partial t}(t,s) + \frac{\partial x}{\partial s}(t,s) &=& - x(t,s) \, \sum_{j=1}^m \frac{g_j(s)}{\zeta_j(t)} \\
\frac{\partial l}{\partial t}(t,s) + \frac{\partial l}{\partial s}(t,s) &=&  
- x(t-h,s-h) \, \sum_{j=1}^m \frac{g_j(s-h)}{\zeta_j(t-h)} \, \theta(s-h) \nonumber
\ee}
with boundary conditions
\be\label{sys1:bc}
x(t,0) = l(t,0) =  1
\ee
and where
\be
\zeta_j(t) = \int_{0}^{t-h} g_j(s) \, l(t,s) \, d s
\ee
These equations are deterministic and describe the mean-field behavior of the 
system. Therefore they provide information about the average behavior of the Tangle in the regime of large arrival rate.
Note that for $t \ge h$   the second equation in (\ref{sys1}) can be solved using (\ref{sys1:bc}):
\be\label{sys1:l-sol}
l(t,s) = \begin{cases} 1 & 0 \le s \le h \\
x(t-h,s-h) & h \le s \le t \end{cases}
\ee
So for $t \ge h$ we will consider the {simpler} equivalent system
\be\label{sys2}
\frac{\partial x}{\partial t}(t,s) + \frac{\partial x}{\partial s}(t,s) = - x(t,s) \, \sum_{j=1}^m \frac{g_j(s)}{\zeta_j(t)} , \qquad
x(t,0) =  1
\ee

\medskip
{\color{black}
We first state and prove an existence result for solutions of the system (\ref{sys2}) for $t \ge 2h$.
The system must be supplemented with suitable initial conditions, 
{and it is convenient to formulate these
in terms of initial functions in the interval $h \le t \le 2h$}.
Accordingly let $\{\phi(s), \psi_j(t)\}$ $(0 \le s \le 2h, \, 0 \le t \le h)$ be continuously differentiable non-negative
functions satisfying the following conditions:
{
\small
\be\label{sys1:ic}
\phi(0) &=& 1 \nonumber \\
\phi'(0) &=& - \sum_{j=1}^m \frac{g_j(0)}{\psi_j(0)} \nonumber \\
\psi_j(0) &=& \int_{0}^h g_j(s) \, ds \nonumber \\
\psi_j(h) &=& \psi_j(0) + \int_{h}^{2h} g_j(s) \, \phi(s-h) \, d s  \nonumber \\
\psi_j(t) & \ge & \psi_j(0) \quad \text{for all $0 \le t \le h$} \nonumber \\
\psi_j'(h) &=& g_j(2h) \phi(h)   \\
 & & - \int_{h}^{2h} g_j(s) \left[\sum_{j=1}^m \frac{g_j(s-h)}{\psi_j(0)} \phi(s-h) + \phi'(s-h) \right] \, ds \nonumber
\ee}

\begin{thm}\label{thm0}
Suppose that $\{g_j\}$ $(j=1,\dots,m)$ are locally integrable non-negative functions on $\mathbb{R}_+$,
with $ \int_{0}^h g_j(s) \, ds > 0$ for all $j=1,\dots,m$.
Let $\{\phi, \psi_j\}$ be continuously differentiable non-negative functions satisfying the conditions (\ref{sys1:ic}).
Then there are unique continuously differentiable functions
$\{\zeta_j(t), \,  x(t,s)\}  \, (t \ge 2h, 0 \le s \le t)$ satisfying
\be\label{thm0:eq1}
x(2h,s) &=&  \phi(s), \quad 0 \le s \le 2 h \\
\zeta_j(t) &=& \psi_j(t - 2h), \quad 2h \le t \le 3h \nonumber
\ee
and such that $x(t,s)$ satisfies (\ref{sys2}) for all $t \ge 2h, 0 \le s \le t$.
\end{thm}

\par\noindent{\em Proof:}
the result is proved by first solving the system (\ref{sys2}) for $x(t,s)$ in the interval $2h \le t \le 3h$, then using this to generate the
functions $\zeta_j(t)$ in the interval $3h \le t \le 4h$, then using these to solve for $x(t,s)$ in the interval
$3h \le t \le 4h$, and so on.
It will be convenient to introduce the function
\be\label{def:f}
f(t,s) = \sum_{j=1}^m \frac{g_j(s)}{\zeta_j(t)}
\ee
Note that by assumption $\zeta_j(2h) > 0$, so $f(2h,s)$ is well-defined.
We will see that $\zeta_j(t) \ge \zeta_j(2h)$ for all $t \ge 2h$, so (\ref{def:f}) is well-defined
for all $t \ge 2h$.
The following definition will play a key role: for $0 \le v \le s \le t$ define
\be\label{def:Prop}
P(t,s,v) & = & \exp \left( - \int_{v}^s \sum_{j=1}^m \frac{g_j(w)}{\zeta_j(t+w - s)} \, d w \right) =\nonumber \\
& = &  \exp \left( - \int_{v}^s f(t+w - s,w) \, d w \right)
\ee
We define $\zeta_j(t) = \psi_j(t-2h)$ for $2h \le t \le 3h$, and $x(2h,s) = \phi(s)$ for $0 \le s \le 2h$. We
then have from (\ref{sys2})
{\small
\be\label{x:2h,3h}
x(t,s) = \begin{cases}
P(t,s,0) & 0 \le s \le t - 2h \\ P(t,s,2h-t+s) \, \phi(2h -t +s) & t-2h \le s \le t \end{cases}
\ee}
This defines $x(t,s)$ for all $2h \le t \le 3h, \, 0 \le s \le t$, and it is the unique solution satisfying the boundary conditions
$x(2h,s) = \phi(s)$ and $x(t,0)=1$. From (\ref{def:Prop}) it is immediately clear that $x$ is continuously differentiable
except possibly on the line $s = t-2h$ where the two cases overlap. Since $\phi(0)=1$
it follows that $x$ is continuous on this line. We can calculate the derivatives for the two cases and find
\be
\frac{\partial x}{\partial s} = 
P \left[ -f(t,s) + \int_{0}^s \partial_1 f(t+w-s,w) \, dw \right]
\ee
for $ 0 \le s \le t - 2h$, and
\be
\frac{\partial x}{\partial s} &=& 
P\biggl[- f(t,s) + \int_{2h-t+s}^s \partial_1 f(t+w-s,w) \, dw  \\
&& + f(2h,2h-t+s) \biggr] \phi(2h-t+s)  + P \phi'(2h-t+s) \nonumber
\ee
for $ t-2h \le s \le t$. Using the initial conditions (\ref{sys1:ic}) it follows that these two expressions are equal when $s= t-2h$,
hence $\frac{\partial x}{\partial s}$ is continuous in the region  $2h \le t \le 3h, \, 0 \le s \le t$.
Since the equation (\ref{sys2}) holds, it follows that also $\frac{\partial x}{\partial t}$ is continuous in the interior of this region.
At the boundary $s=0$ we have the boundary condition $x(t,0)=1$ so we must check that $\frac{\partial x}{\partial t} = 0$
at $s=0$. This involves only the first expression on the right side of (\ref{x:2h,3h}) and it is easy to verify that the derivative does vanish
at $s=0$.

\medskip
We now use this function to generate $\zeta_j$ in the interval $3h \le t \le 4h$. First we use (\ref{sys1:l-sol})  to compute $l(t,s)$,
then we find from (\ref{def:zeta})
\be\label{compute_zeta1}
\zeta_j(t) = \int_{0}^h g_j(s) \, ds + \int_{h}^{t-h} g_j(s) x(t-h,s-h) \, d s
\ee
We use (\ref{sys1:ic}) to verify that  $\zeta_j(3h) = \psi_j(h)$,
so $\zeta_j$ is continuous at $t=3h$.
Furthermore we can compute the derivative of (\ref{compute_zeta1}) at $t=3h$, and check that it matches the value
$\psi_j'(h)$ specified in (\ref{sys1:ic}). Therefore $\zeta_j(t)$ is defined and continuously differentiable in the interval
$2h \le t \le 4h$. We now continue the bootstrap and use this to compute $x(t,s)$ for $3h \le t \le 4h$:
{
\small
\be\label{x:3h,4h}
x(t,s) \hspace{-3 mm}&=& \hspace{-3 mm}
\begin{cases}
P(t,s,0) & 0 \le s \le t - 3h \\ P(t,s,3h-t+s) x(3h, 3h -t +s) & t-3h \le s \le t \end{cases} \nonumber \\
&\hspace{-3 mm}=&\hspace{-3 mm} \nonumber
\begin{cases}
P(t,s,0) & 0 \le s \le t - 2h \\ P(t,s,2h-t+s) \phi(2h -t +s) & t-2h \le s \le t \end{cases}
\ee}
As before we can show that $x(t,s)$ is continuously differentiable in the region
$3h \le t \le 4h, 0 \le s \le t$, and satisfies the equations (\ref{sys2}).
This allows us again to define $l$ and then compute $\zeta_j$ in the interval $4h \le t \le 5h$:
\be\label{compute_zeta2}
\zeta_j(t) = \int_{0}^h g_j(s) \, ds + \int_{h}^{t-h} g_j(s) x(t-h,s-h) \, d s
\ee
The construction can be continued, generating the solution $x(t,s)$ for all $t \ge 2h$, $0 \le s \le t$ as claimed.}

\subsection{Integrable weights: persistence of orphans}
In this section we consider the case where all the weight functions are integrable, that is
\be\label{g-int-all}
\int_{0}^{\infty} g_j(s) \, d s < \infty, \quad j=1,\dots,m
\ee
For example this situation is expected to hold when tips are selected using the MCMC algorithm {with a sufficiently large value of $\alpha$, as our simulations in a previous section show}. 
\textcolor{black}{We will not consider here the general question of the stability of solutions of
the system (\ref{sys1}) as $t \rightarrow \infty$ (although Theorem \ref{thm2}
does provide some information about this). Rather, in Theorem \ref{thm1} we are interested in exploring and characterizing the steady state behaviour
which has been observed in simulations of the Tangle, by us and by the IOTA foundation \cite{simulations}.
In order to describe this steady state behaviour, we derive a time-independent version of the system (\ref{sys1}),
see (\ref{sys3}) below.
In Theorem \ref{thm1} we state and prove some properties of this system (\ref{sys3}) pertaining to the persistence of orphaned transactions.
We will show that for this system the total number of unapproved tips
is infinite. 
If the Tangle converges to an equilibrium average behavior, as suggested by simulations, then this result would support the conclusion that 
orphans must persist, meaning that the set
of unapproved  tips must grow without bound. 
Thus the results of Theorem \ref{thm1} provide a strong argument in favor of developing
new tip selection algorithms which can address the problem of persistence of orphaned transactions.
The steady state solutions $\{l(s), \, x(s)\}$ satisfy the following system,
which is derived from (\ref{sys1}) by replacing $\{l(t,s), \, x(t,s)\}$ by $\{l(s), \, x(s)\}$ and letting $t \rightarrow \infty$:
\be\label{sys3}
\frac{d x}{d s}(s) &=& - x(s) \, \sum_{j=1}^m \frac{g_j(s)}{\zeta_j} \nonumber \\
\frac{d l}{d s}(s) &=& - x(s-h) \, \sum_{j=1}^m \frac{g_j(s-h)}{\zeta_j} \, \theta(s-h) \nonumber \\
x(0) &=& 1 \nonumber  \\
l(0) &=& 1  \nonumber \\
\zeta_j &=& \int_{0}^{\infty} g_j(s) \, l(s) \, d s
\ee
}

\textcolor{black}{\begin{thm}\label{thm1}
Suppose (\ref{g-int-all}) holds. Then any solutions $l(s), \, x(s)$ of the system (\ref{sys3}) satisfy
\be
\int_{0}^{\infty} l(s) ds = \int_{0}^{\infty} x(s) ds = \infty
\ee
\end{thm}}

\par\noindent{\em Proof:}
It follows immediately that $l(s) = l(0) = 1$ for $0 \le s \le h$, and that
\be
l(s) = x(s-h) \quad \mbox{for $s \ge h$}
\ee
Furthermore $l(s)$ is decreasing and so
\be
\zeta_j \le \int_{0}^{\infty} g_j(s) \, l(0)  \, d s < \infty \quad j=1,\dots,m
\ee
The solution of (\ref{sys3}) is
\be
x(s) = \exp \left[ - \sum_{j=1}^m \, \zeta_j^{-1} \, \int_{0}^{s} g_j(u) d u \right]
\ee
and thus it follows that
\be
\lim_{s \rightarrow \infty} x(s) = \exp \left[ - \sum_{j=1}^m \, \zeta_j^{-1} \, \int_{0}^{\infty} g_j(u) d u \right] > 0
\ee
Therefore
\be
\lim_{s \rightarrow \infty} \int_{0}^{s} x(u) \, d u = \infty
\ee
Since $l(s) \ge x(s)$ this also implies that $l(s)$ is not integrable.\newline

\emph{Remark:} as noted before, Theorem \ref{thm1} supports the conclusion that when all tip selection algorithms are
'short-range', meaning that all selections favor more recently arrived transactions, the number of orphan 
transactions must grow without bound.

\subsection{Random tip selection: eventual approval of all transactions}
We now consider the situation where at least two of the tips are selected randomly, meaning that the
corresponding weight functions are constant. Without loss of generality these can be the first two
tip selections, so we assume
\be\label{non-int1}
g_1(s) = g_2(s) = 1
\ee
\textcolor{black}{
We state and prove a Theorem which shows that in this situation the number of unapproved tips
remains bounded as the Tangle grows.}

\textcolor{black}{\begin{thm}\label{thm2}
Let $\{x(t,s), l(t,s)\}$ be a solution of the system (\ref{sys1}) as described in Theorem \ref{thm0}, and assume that {(\ref{bound-w}) and (\ref{non-int1}) hold}.
Then $l(t)$ is uniformly bounded for all $t \ge 0$. 
\end{thm}}

\par\noindent{\em Proof:}
applying (\ref{non-int1}) to  (\ref{sys1}) we get for $t \ge 2 h$
\be\label{ineq4}
\frac{\partial x}{\partial t}(t,s) + \frac{\partial x}{\partial s}(t,s) \le - \frac{2 \, x(t,s)}{l(t)}
\ee
Recall that
\be
x(t) = \int_{0}^{t-h} x(t,s) d s
\ee
and therefore
\be
\frac{d x}{d t} = x(t,t-h) + \int_{0}^{t-h} \frac{\partial x}{\partial t}(t,s) \, d s
\ee
Also we have
\be
\int_{0}^{t-h} \frac{\partial x}{\partial s}(t,s) \, d s = x(t,t-h) - x(t,0) = x(t,t-h) - 1
\ee
Therefore integrating over $s$ in the inequality (\ref{ineq4}) we get
\be\label{ineq5}
\frac{d x}{d t} \le 1 - 2 \, \frac{x(t)}{l(t)}
\ee
Furthermore we have $l(t) = x(t) + w(t)$ and thus from (\ref{bound-w}) we derive
\be
l(t) \le x(t) + m h
\ee
Substituting into (\ref{ineq5}), it gives
\be\label{ineq6}
\frac{d x}{d t} \le 1 - 2 \, \frac{x(t)}{x(t) + m h}
\ee
It follows from (\ref{ineq6}) that 
\be\label{ineq7}
\frac{d x}{d t}(t) < 0 \quad \mbox{for $x(t) > m h$}
\ee
\textcolor{black}{
Recall that the value $x(2h)$ is set by the initial conditions (\ref{thm0:eq1}).
Therefore from (\ref{ineq7}) we deduce
\be
\sup_{t \ge 0} x(t) \le \max \{m h, x(2h) \}
\ee
Since $l(t) \le x(t) + m h$ this proves that $l(t)$ is uniformly bounded.
}

\emph{Remark:} assuming that the distribution of tips approaches an equilibrium as $t \rightarrow \infty$, 
the result that $l(t)$ is uniformly bounded means that the density of tips at age $s$ converges to zero as $s \rightarrow \infty$.
This means that, except possibly for a transient set of tips created by the initial conditions,
all tips will be eventually approved under the assumptions of Theorem \ref{thm2}.
We conjecture that the same result holds under the weaker assumption that at least one of the tips is randomly selected.
This conjecture would imply that $l(t)$ is uniformly bounded for our hybrid algorithm where one tip is selected randomly,
and the other tip is selected by the MCMC algorithm. Our simulations of the Tangle presented  in Section \ref{sec: Tangle} support this conjecture.
Notice that in these simulations the number of tips $L(t)$ diverges or converges depending on the 
size of the parameter $\alpha$ (in the proposed model changing $\alpha$ is equivalent to changing the function $g(\cdot)$: larger $\alpha$ is closer to MCMC, smaller $\alpha$ is closer to the random tip selection model). This supports our contention that the PDE model is consistent and matches the behavior of the agent based model of the Tangle.

\section{Conclusions}
\label{sec: Conclusions and Future Research}
\textcolor{black}{Directed Acylic Graphs (DAGs)  are emerging as an attractive alternative to traditional blockchain  architectures 
for distributed ledger technology (DLT). 
Our principal contribution is to propose a simple modification to the attachment mechanism 
for the Tangle (the IOTA DAG architecture). This modification ensures
that  all transactions 
are validated in finite time, and preserves essential features of the popular Monte-Carlo selection algorithm. A fluid approximation for the Tangle is also derived and shown to
exhibits desired behavior. Finally, we also present simulations which validate the results for finite arrival rates.}
 \textcolor{black}{As a final remark we note that the hybrid algorithm presented in this paper is being evaluated and tested by the IOTA foundation: see text in  \cite{coordicide} that refers in reference [21] to this paper. Note that [21] is the ArXiv version of this present paper.}

\section{Acknowledgements}
The work is partly supported by the Danish ForskEL programme (now EUDP) through the Energy Collective project (grant no. 2016- 1-12530) and by SFI grant 16/IA/4610.

\begin{biography}
  [{\includegraphics[width=1in,height=1.25in,clip,keepaspectratio]{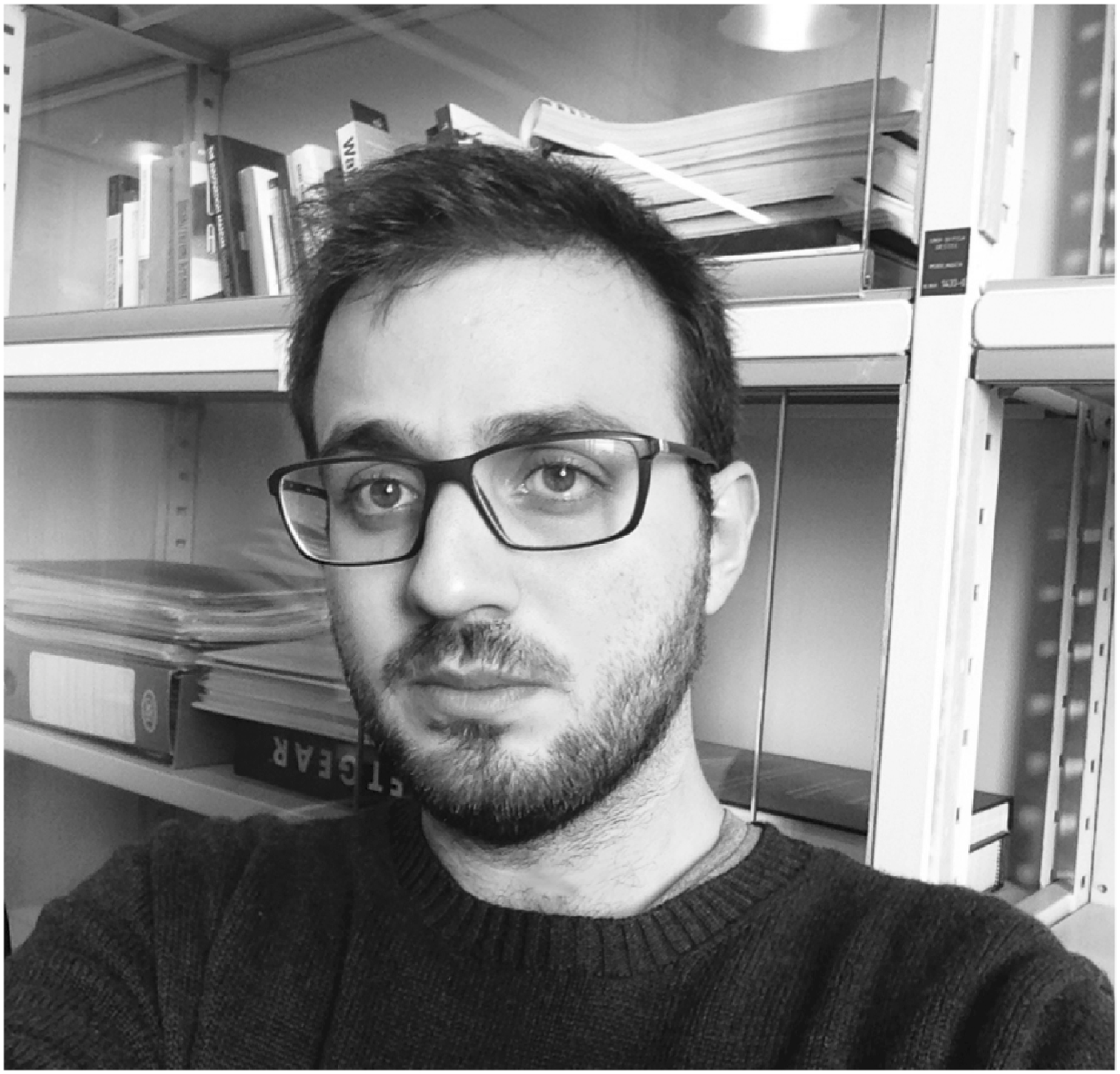}}]
  {Pietro Ferraro} received a PhD in control and electrical engineering from the University of Pisa, Italy, in 2018. He is currently a Post Doc Fellow with the school of electrical and electronic engineering at University College Dublin (UCD). His research interests include control theory applied to the sharing economy domain.
\end{biography}
\vskip -2\baselineskip plus -1fil
\begin{biography}[{\includegraphics[width=1in,height=1.25in,clip,keepaspectratio]{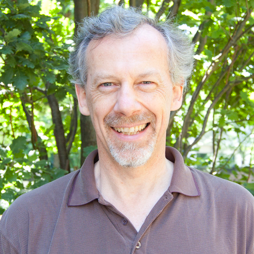}}]
  {Christopher King} is currently a Professor in the Mathematics Department at Northeastern University. His research interests include dynamical systems, quantum information theory, and mathematical physics.

\end{biography}
\vskip -2\baselineskip plus -1fil

\begin{biography}[{\includegraphics[width=1in,height=1.25in,clip,keepaspectratio]{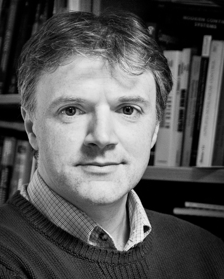}}]
  {Robert Shorten} is Professor of Control Engineering and Decision Science at UCD. He was a co-founder of the Hamilton Institute at Maynooth University, and led the Optimisation and Control team at IBM Research Smart Cities Research Lab in Dublin Ireland. He has been a visiting professor at TU Berlin and a research visitor at Yale University and Technion. He is the Irish member of the European Union Control Association assembly, a member of the IEEE Control Systems Society Technical Group on Smart Cities, and a member of the IFAC Technical Committees for Automotive Control, and for Discrete Event and Hybrid Systems. He is a co-author of the recently published books AIMD Dynamics and Distributed Resource Allocation (SIAM 2016) and Electric and Plug-in Vehicle Networks: Optimisation and Control (CRC Press, Taylor and Francis Group, 2017)
\end{biography}

\end{document}